# Quantum dephasing of kagome superconductivity

Jia-Xin Yin

Department of Physics, Southern University of Science and Technology, Shenzhen, Guangdong 518055, China.

Quantum Science Center of Guangdong-Hong Kong-Macao Greater Bay Area (Guangdong), Shenzhen 518045, China

E-mail: yinjx@sustech.edu.cn

**Recent observations for a pressurized kagome superconductor through transport, muon spin relaxation, nuclear magnetic resonance, and point contact spectroscopy demonstrate striking fluctuating superconductivity at a magic pressure. This discovery may hint at a new quantum dephasing mechanism, which can be discussed from the perspectives of correlation and topology. Three outstanding questions should be considered in decoding this mechanism: one is the intrinsic phase fragile nature of kagome superconductivity; second is the role of Coulomb interactions within different charge-ordered phases; third is the geometrical phase compatibility between superconductivity and different charge orders.**

Researchers have been fascinated by the complex phase diagrams for certain quantum materials for decades. One prime example is the phase diagram for cuprate high-temperature superconductors, where many researchers believe that the many-body interactions leading to its rich phase diagram can eventually unveil the receipt for high-temperature superconductivity (SC) in quantum matter. While it does not exhibit high-temperature SC, the recently discovered $CsV_3Sb_5$ family of kagome superconductors has surprised the community through its unconventional high-temperature charge density wave order (CDW) and unusual low-temperature SC behaviors. Two research themes are outlined, including correlation effects and topological matter [1]. One intriguing discovery for correlation effects is about the two SC domes in the temperature-pressure phase diagram of $CsV_3Sb_5$ [Fig. 1(a)] [2,3].

When the double SC dome is discovered, it follows with another puzzling observation that the CDW above the two SC domes seems to be quite different by simply looking at the resistivity kink behavior at the CDW transition. To resolve their difference as well as their relationship with SC, advanced spectroscopic techniques are required to look into a pressurized crystal, which is extremely challenging. Three spectroscopic techniques under high pressure have been leveraged, including muon spin relaxation (μSR) [4], nuclear magnetic resonance (NMR) [5], and point contact spectroscopy (PCS) [6][7]. A key finding of the μSR experiment is the observation of two superfluid-density domes in scale with SC [4]. The NMR experiment identifies substantial spectroscopic differences between the two types of CDW. Despite lacking direct wavelength resolution, a stripe-like 1 × 4 CDW above the double dome dip has been proposed to explain the strong lineshape anomaly of the NMR data [5], which is in addition to the widely accepted 2 × 2 CDW phase at ambient pressure [8]. Now, two research teams further use PCS to add a new chapter to the story [6][7]: they independently report the emergence of larger superconducting gaps at the magic pressure between the two SC domes. This discovery together with previous findings makes the whole phase diagram including superfluid density and pairing strength identical to that of cuprates $(La,Ba)CuO_4$, and highlights strongly fluctuating kagome superconductivity at the magic pressure. In cuprates $(La,Ba)CuO_4$, it has long been discussed that at the magic 1/8 hole doping, charges localize into 1 × 4 stripes in an antiferromagnetic background, and the stripes strongly suppress SC but not the pairing strength of the *d*-wave superconductivity. The comparison between the two-dome kagome SC and two-dome cuprates SC at this point is striking, yet the underlying mechanism of fluctuating kagome superconductivity is far from clear.

Superconductivity is a macroscopic effect caused by a condensation of Cooper pairs into a boson-like state. Its microscopic order parameter is a complex number $\Delta e^{i\theta}$, which includes the pairing energy $\Delta$ and pairing phase $\theta$ [8]. In conventional superconductors described by the Bardeen-Cooper-Schrieffer (BCS) theory, once electrons are paired, they concurrently condensate with a global phase coherence, and their SC transition temperature $T_C$ is governed by the $\Delta$. However, for superconductors with low superfluid density, the phase ordering temperature can be lower than the pairing ordering temperature due to phase fluctuations, and consequently, the phase coherence plays a more important role in determining $T_C$ [Fig. 1(b)]. Research in cuprate high-temperature superconductivity that exhibits small superfluid density often discusses the concepts of the pseudogap, preformed pairs, and phase fluctuation. In this context, one may conclude that the µSR experiment (scaling between $T_C$ and superfluid density) and PCS experiment (enhanced $\Delta$ in opposite to $T_C$ between the two domes) taken together tell us the existence of strong phase fluctuations at the magic pressure. Then an outstanding question would be how does the $1 \times 4$ CDW at the magic pressure uniquely cause such strong phase fluctuation? We may look at this question from the perspectives of correlation and topology.

From the perspective of correlation, one generically discussed key factor for phase fluctuation is the poorly screened Coulomb interaction. As Coulomb interaction between charges reduces their number fluctuations, which promotes phase fluctuations according to the number-phase uncertainty rule [9]. The strong Coulomb interaction naturally arises for cuprates as their parent compounds are Mott insulators. On the superconductivity side, µSR experiments have found that the kagome superfluid density is very low [Fig. 1(c)][1,4], close to that of cuprates, indicative of a stronger potential for exhibiting phase fluctuations. The phase's fragile nature would suggest the existence of certain poorly screened Coulomb interactions in $CsV_3Sb_5$. The on-site Coulomb interaction U for $CsV_3Sb_5$ is evaluated to be weak by comparing electronic bands from experiments and theories [1]. On the other hand, the extended Coulomb interaction V is often proposed to be accountable for the emergence of time-reversal symmetry-breaking CDW discussed in $CsV_3Sb_5$ family of materials [1,4,6,10]. Therefore, the role of U and V in determining the phase fragile nature of kagome superconductivity should be the one outstanding question.

On the CDW side, $CsV_3Sb_5$ seems do not have a strong antiferromagnetic spin order as a background for the formation of charge localization as stripes in cuprates. Experimentally, $1 \times 4$ modulations have been detected in $CsV_3Sb_5$ by surface techniques such as scanning tunneling microscopy but are not yet confirmed by bulk scattering techniques such as neutron scattering and X-ray scattering, thus is often speculated as a surface phenomenon supported with first-principles calculations. Inspired by these intensive discussions that are not covered here, a pioneering first-principles calculation does find the $1 \times 4$ CDW as a meta-stable phase in addition to the $2 \times 2$ CDW within the kagome lattice [Fig. 1 (d)][11]. This type of $1 \times 4$ CDW is driven by the dimer formation of V atoms plus an antiphase coupling between kagome layers, thus is of structural origin rather than spin driven. Such $1 \times 4$ CDW instability disappears with reduced chemical pressure [11], thus is expected to be more stable with adding physical pressure, serving as a candidate for the pressure-induced CDW. While the absolute Coulomb interaction may not change much between the two CDW phases (as inferred from the NMR results on the universal scaling of relaxation rates[5]), from their charge distribution in Fig. 1(d), the inter-distance between charge clusters for $1 \times 4$ CDW is larger (owing to much-reduced charge on certain sites), which reduces the effective electron hopping energy *t* [12] relative to the Coulomb energy. Then whether the role of Coulomb interaction (U/t or V/t) in the $1 \times 4$ CDW can be enhanced in this way deserves further investigation, which may eventually deepen our understanding of the phase fluctuations.

From the viewpoint of topology, while cuprates are generally believed to exhibit a *d*-wave pairing state, the SC for $CsV_3Sb_5$ family of materials is recently reported to be nodeless by high-resolution angle-resolved photoemission spectroscopy [Fig. 2(a)][13], as well as to be time-reversal symmetry-breaking by muon experiments [Fig. 2(b)][4,13,14]. One promising candidate for satisfying these two constraints is the chiral d+id pairing [Fig. 2(c)], which has been considered for kagome superconductivity at the van Hove filling [15] even before the $CsV_3Sb_5$ superconductor is discovered. A kagome lattice has three sublattices, and d+id pairing can be derived by considering pairing between nearest neighbor sites within each sublattice, with further exhibiting a $2\pi/3$ winding phase between three degenerate lattice directions. It is conceivable that if this pairing symmetry is correct, the phase fluctuations would be associated with disturbing the global phase coherence of Cooper pairs with each carrying a winding phase. Interestingly, when the $2 \times 2$ CDW is discovered, the initially discussed time-reversal symmetry-breaking CDW order parameter (often termed as chiral charge order or topological charge order) also has a similar $2\pi/3$ winding phase between three degenerate kagome lattice directions [Fig. 2(d)][6]. In connection with the correlation discussions, both the chiral charge order and chiral SC require substantial Coulomb interactions to become the ground state. And both the chiral charge order and chiral SC can carry several topological properties, including orbital currents, orbital magnetization, and chiral edge modes. This situation means that the quantum phase of $2 \times 2$ CDW can be highly compatible with that of SC [Fig. 2 (e)]. Meanwhile, it is hard to imagine that the $1 \times 4$ CDW can have a well-defined winding phase. Whether such geometrical phase incompatibility background can be related to enhanced phase fluctuations also deserves further investigation with new theoretical approaches [16].

Therefore, these new observations not only show a striking analogy between kagome SC and cuprates SC, but also seem to be related to other frontier topics in kagome superconductivity, including weak superfluid density, multiple CDW vectors, time-reversal symmetry-breaking CDW and chiral SC. These advanced spectroscopic experiments (μSR, NMR, and PCS) under pressure taken together may provide a rare example of the dephasing of superconductivity from the intertwining of correlation and topology in quantum matter. We look forward to more experimental and theoretical advances in elucidating the exact quantum mechanisms of the fluctuating kagome superconductivity.

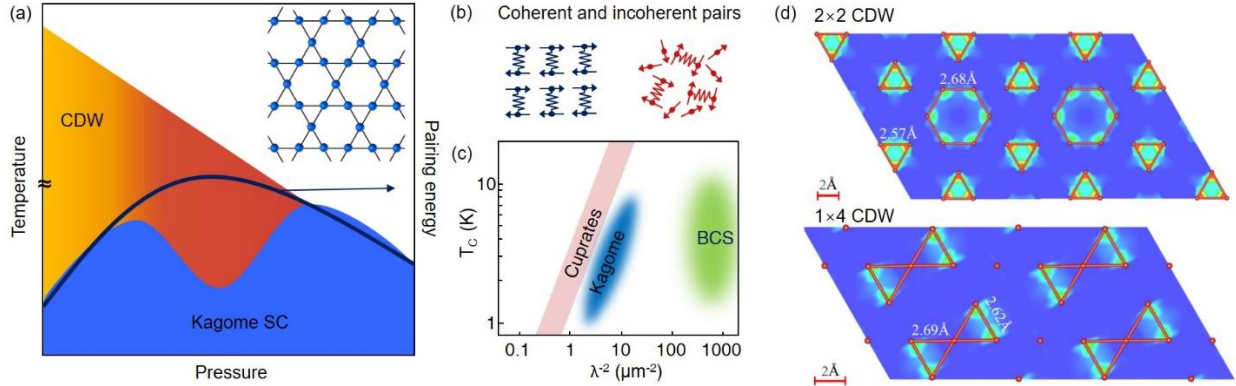

**Fig. 1** (a) Schematic phase diagram of kagome superconductor $CsV_3Sb_5$ under pressure. With increasing pressure, the CDW becomes progressively suppressed and evolves from one type of CDW (gold color) to another (orange color). The kagome superconductivity (SC) exhibits a striking two domes feature while the electron pairing energy exhibits a single-dome feature. (b) Schematic of coherent Cooper pairs and incoherent Cooper pairs. (c) Scaling plot for superconducting transition temperature $T_C$ versus inverse squared magnetic penetration depth. The scaling relation of kagome superconductors is close to that of cuprates but away from that of conventional Bardeen-Cooper-Schrieffer (BCS) superconductors. The thin superfluid density of kagome superconductors makes them have a stronger potential to exhibit phase fluctuations. (d) A pioneering first-principles result for two types of CDW within the kagome lattice for $CsV_3Sb_5$.

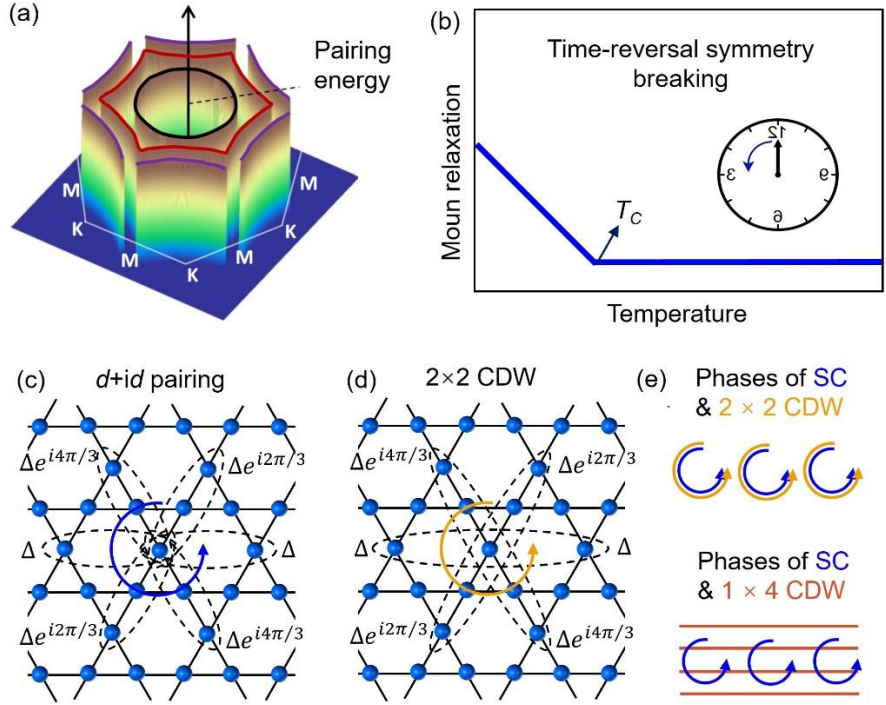

**Fig. 2** (a) Illustration of electron pairing amplitude in momentum space as detected by angle resolved photoemission for $CsV_3Sb_5$ family of kagome superconductor, essentially showing a nodeless superconducting gap structure. (b) Illustration of time-reversal symmetry breaking superconducting state detected by muon rotation spectroscopy for $CsV_3Sb_5$ family of kagome superconductor. (c) A candidate pairing state d+id as initially considered for kagome superconductivity at the van Hove filling which is proposed before the discovery of $CsV_3Sb_5$. This pairing state is consistent with the nodeless gap and time-reversal symmetry-breaking features. This chiral pairing state uniquely features a winding phase. (d) A candidate 2×2 CDW as initially considered in experimental observation of the CDW phase in kagome superconductors. This CDW phase also breaks time-reversal symmetry and uniquely features a similar winding phase. (e) Schematic showing that the initially considered 2 × 2 CDW also exhibits a winding phase that matches with the chiral superconductivity, while the 1 × 4 CDW could not exhibit a match with the chiral superconducting phase.